\documentclass[twocolumn,showpacs,preprintnumbers,amsmath,amssymb]{revtex4}
\usepackage{graphicx}% Include figure files
\usepackage{dcolumn}% Align table columns on decimal point
\usepackage{bm}% bold math

\begin{document}

\preprint{APS/123-QED}

\title{Conductivity, weak ferromagnetism and charge instability in $\alpha-  MnS $ single
crystal }
\author{S.S. Aplesnin , L. I. Ryabinkina ,G. M. Abramova,
O.B. Romanova, A.M. Vorotynov, D.A. Velikanov,N. I. Kiselev ,  A.
D. Balaev}

\affiliation{%
 Kirensky Institute of Physics, Siberian Branch of the
Russian Academy of Sciences, Krasnoyarsk, 660036, Russia\\
%This line break forced with \textbackslash\textbackslash
}%
 \email{apl@iph.krasn.ru}
\date{\today}% It is always \today, today,
             %  but any date may be explicitly specified

\begin{abstract}

The temperature  dependence of resistivity , magnetization and
electron-spin resonance of the $\alpha-  MnS $ single crystal were
measured in temperature range of $5 K < T < 550 K$. Magnetization
hysteresis  in applied magnetic field up to 0.7 T at $T=5 K, 77 K,
300 K$, irreversible temperature behavior of magnetization and
resistivity  were found . The obtained data were explained in
terms of degenerate tight binding  model using random phase
approximation. The contribution of holes in $t_{2g}$ and $e_g$
bands of manganese ions to the conductivity,  optical absorbtion
spectra and  charge instability  in  $\alpha -MnS $ were studied.
Charge susceptibility maxima resulted from the competition of the
on-site Coulomb interaction between the holes in different
orbitals and small hybridization of sub-bands were calculated at
$T=160 K, 250 K, 475 K$.

\end{abstract}

% PACS, the Physics and Astronomy
                             % Classification Scheme.
%\keywords{Suggested keywords}%Use showkeys class option if keyword
                              %display desired
\maketitle

%\section{\label{sec:level1}First-level heading:\protect\\
% \textbackslash\textbackslash
 {\bf I. Introduction}\\

Recently the compounds with strong coupling of the charge-orbital
and spin degrees of freedom have attracted extensive interest in
connection with the specific  property, namely the colossal
magnetoresistance, i.e. the strong resistivity decrease in applied
magnetic field. In manganites with small concentration of doped
carriers this strong coupling leads to ferromagnetism, which
results from the competition between ferromagnetic double-exchange
interaction and  AFM super-exchange. High temperature
 behavior of the resistivity $\rho(T)$ in the manganites $RMnO_3 (R=La, Pr, Sm) $ \cite{Manganit}
  looks  like  step  function explained by the charge  and orbital ordering. Since the substitution of manganese ion
  by another
 3d-metal ion leads to the absence of these properties
, the dramatic resistivity drop in applied magnetic field is
attributed to the electronic  state of the
manganese ion. \\
 Similar effects   have also been observed in compounds
  synthesized on basis of $\alpha-MnS$. The diluted magnetic semiconductors
$Mn_{(1-x)}Fe_xS$ reveal the colossal magnetoresistance.
 The properties of these compounds should be caused by the electron structure of the manganese monosulfide, which
results in a number of characteristic properties of $MnS$ pure
single crystals. Similar with $LaMnO_3, \alpha-MnS $ sulfide shows
antiferromagnetic $(AF)$ ordering of the second kind consisted of
ferromagnetically arranged spins in $(111)$ plane and AF spin
ordering along cube edges .   Unlike $LaMnO_3$ with $Mn^{3+}$
ions, the ground state  of the manganese ions in $MnS$ with
$NaCl-$ structure is $Mn^{2+}$. The resistivity of the $MnS$ pure
single crystals  is independent of temperature   at $T<T_N$
\cite{resist} and behaves analogously to semiconductors up to $800
K$ . In temperature range of $400 K-550K $  the resistivity of the
manganese monosulfide has a plateau \cite{Heikens} , the mechanism
of which is not studied yet.  As the Hall effect measurements
\cite{Heikens}showed, the conductivity is realized by holes in
3d-band of
manganese ions.\\
 The electronic and magnetic
properties of $ \alpha-MnS $ have been studied in the framework of
the density functional level theory by  self-consistent solving
the Kohn-Sham equation  \cite{Taperro}. First principle
calculations confirm the hole character of the conductivity. The
$e_g$ and $t_{2g}$  bandwidths of $Mn$ states are $\sim 2.5 eV $
and $\sim 1 eV$, respectively. The $t_{2g}$ bands corresponding to
the spin-up and spin-down electron states are separated by $\sim
1.6 eV$.
 Fermi level is located at the bottom of the $t_{2g}$ band  with the spin down. The states of
the valence band are occupied by electrons of both $p$ sulfur and
$d$ manganese orbitals  and the
gap value is $\sim 1.5 eV$.\\
In this paper we determine the contribution of the carriers in the
upper $e_g$ and $t_{2g} $ Hubbard  bands to the conductivity at
low and high temperatures and the influence of charge instability
on the transport and magnetic properties of $ \alpha-MnS $. In
contrast to manganite, the weak magnetic moment results from
 polarization of the holes spins localized in one of the
subbands $t_{2g} $ by spin-orbital interaction.\\

{\bf 2. Experimental results}\\

The $\alpha - MnS$ single crystal was made by liquid manganese
saturation with sulfur at $T \sim 1245 ^0 C $. X-ray diffraction
analysis was performed on $DRON-2.0$ diffractometer with the
monochromatic $ CuK \alpha- $ radiation at temperatures $80 - 300
K $.  The resistance measurements were made in $[111]$ and $[100]$
directions  at  temperatures  $90- 550 K$. The fluorescence
spectroscopy experiment was carried out on SPARK-1 spectrometer.
According to the X-ray analysis data, synthesized $\alpha - MnS$
sample is the single crystal, which has  $NaCl$ cubic
lattice with the cell parameter $a=5.22 A $.\\
In fig.1 the temperature dependence of the resistivity $\rho$ for
$ \alpha-MnS $  single crystal  is shown. One can clearly see a
gradual change of the resistivity during the sample heating
several times up to 550 K. The first heating-and-cooling cycle
causes significant temperature hysteresis $\Delta R/ R \sim 1$ and
the second cycle yields $\Delta R/ R \sim 0.2 $ at $T=280 K$  .
During the subsequent heating-and-cooling cycles the temperature
hysteresis of the resistivity disappears, and the resistivity has
a plateau  in the range of $420-550K$ showing good agreement with
the previously obtained data \cite{Heikens}. These measurements
were made at high vacuum and the results appeared reproducible
after long
keeping at room temperature. \\
The current - voltage characteristic shows the small negative
differential conductivity at $ U=2 V, 30 V$ and $ T=280 K $, which
disappears at $T=550 K$. The $dI/dU$ curves are presented in
Fig.2. The behavior of $\rho(U) $ at $T=560 K$ is typical for
semiconductors. The magnetic susceptibility measured in range of
$77 K < T < 300 K$ is nonlinear at low magnetic fields  $H=200 Oe,
1000 Oe$ with  anisotropy value $\Delta \chi/\chi$  along $[100]$
and $[111]$ directions  of $\sim 0.16$, and
 correlates with the anisotropy resistivity  $(1- R_{[100]}/ R_{[111]} ) \sim 0.46
 $ \cite{resist}.\\
 Magnetization measurements were carried out with the superconducting
 quantum interference device (SQUID magnetometer). Specimens were cooled
 to $5 K$ and then heated to the highest  temperature  in zero magnetic field .
 The monocrystal was found to have a small spontaneous moment in
 the range of $ 4.2K - 300 K$.  The $m(T)$curves  for  $[001]$ and $[111]$ are shown in Fig.3. After cooling of $\alpha-MnS$ in magnetic field $H=200 Oe $ magnetization $m_{[111]}(T)$ exhibits
 qualitatively different temperature dependence. Magnetization vs $H$
 curves  at $T=5 K, 77 K, 300 K$ are presented in  Fig.4.\\
Electronic spin resonance $(ESR)$ measurements were performed with
the X-band Radiopan SE/X-2544 spectrometer at $\nu \simeq 9.4  GHz
(150 K < T < 300 K)$, using a continuous gas-flow cryostat for $
N_2$. The oriented sample was placed into a quartz tube. Fig.5
shows the temperature dependences of the ESR linewidth $\Delta H$
and the effective $g-$ value $g_{eff}=h\nu/(\mu_B H_{res})$
determined from $H_{res}$. The largest difference between the
linewidths $\delta H= \Delta H_{[001]}-\Delta H_{[100]}$ along
$[001]$ and $[100]$ is observed near the rhombohedral lattice
deformation. The $\delta H$ value changes a sign at $T \simeq 250
K $. The $g-$ values show a small increase at $T >200 K$ and
approach high-temperature value. The zero- field splitting
parameters $(D$ and $E$ is the axial and rhombic terms of the
single ion anisotropy  ) can be determined from temperature
dependence of $H_{res}$ using the general formula for the
resonance shift resulted from the crystal field as will be made
below.

{\bf 3. Model and  calculation method}\\
 The covalent bond between
sulphur and manganese ions leads to redistribution of the electron
density on these ions. The part of sulphur electrons locate on the
$Mn^{2+} $ ion. An addition of $d^6 (t^4_{2g} e^2_g)$ term to main
term of $d^5$ corresponds to partial filling of upper Hubbard
band. Schematic image of the electron density of states of
$Mn^{2+}$ calculated by Taperro et al. \cite{Taperro} is shown in
Fig.6. Fermi level locates below the chemical potential at the
bottom of upper Hubbard band. We neglect the $e_g$ and $t_g$ bands
hybridization, and thus, the conductivity can be estimated as an
additive quantity of $\sigma=\sigma_{e_g}+\sigma_{t_{2g}}$ .
Schematic representation of charge transport through the lattice
sites is shown in Fig.6. Electrostatic interaction of excess
charge lifts the double- and triple degeneration of the $e_g$ and
$ t_{2g}$ sub- bands in the cubic crystal and causes the
rhombohedral deformation of the lattice with the local and global
symmetry breaking.\\
 Theoretical analysis was made in
terms of the degenerate tight binding model with three interaction
parameters: the direct Coulomb integral between the different
orbital electrons $U$, the intraband and interband hopping matrix
elements. Interorbital exchange $J$ is much less than Coulomb
interaction $J/U \sim 0.2 $ and may be omitted.  The motion of the
charge carriers is considered in the paramagnetic phase with the
small holes concentration. This concentration can be estimated
from bandwidth value $W_{t} \sim 1 eV$ and charge transfer gap
calculated from well-known relation
$E_g=\epsilon_p-\epsilon_d+U_{dd}$, where $\epsilon_{p,d}-$ atomic
orbital eigenvalue on the levels $p,d$ for sulphur and  for
manganese ions, $U_{dd}-$ is the intraatomic Coulomb parameter
within the same orbital. Coulomb integral is estimated from
difference of terms energies $E^{+}, E^{3+}$ for the $d^6
(4t_{2g}2e_g)$ and $d^4 (2t_{2g}2e_g)$. The relations
$\Omega_1=E^{3+}-E^{2+}$ and $\Omega_2=E^{+}-E^{2+}$ are found
from X-ray photoelectron spectra \cite{Heikens}  and are equal to
$\Omega_1 \sim - 1.5 eV, \Omega_2 \sim 1.5 eV $ and
$U_{dd}=\Omega_2-\Omega_1 \sim 3 eV $.  The value
$\epsilon_p-\epsilon_d \sim 1 eV $ is equal to the difference of
energies corresponding to maximum of DOS on $3P$ sulphur and $d^5
t_{2g}$ manganese ions \cite{Taperro}.   The concentration is
proportional to the hybridization degree
 $n\sim  (W_t/z E_g)^2 $,which
 is equal to  $n \simeq 5
\cdot 10^{-3}$ and $ 0.01$ for $t_{2g}$ and $e_g$ states,
respectively  . The first principle
 calculations \cite{Taperro} give the higher value of the concentration $n
(t_{2g}) \simeq 0.03, n (e_g) \simeq 0.07$. Appearance of two
electrons on d- level  causes the reduction of the manganese
ion spin $S=4.4 \mu_B$ \cite{Taperro}. \\
Below we consider the motion of charged carriers only in the upper
Hubbard band in the frame of the effective model with the spinless
fermions.    The model Hamiltonian is described as follows:

\begin{eqnarray}
H_t=-\sum_{i,j,\alpha} t_{i,j}^{\alpha \alpha} a_{i\alpha}^+
a_{j\alpha} -\sum_{i,j,\alpha> \beta} t_{i,j}^{\alpha \beta}
a_{i\alpha}^+ a_{j\beta}\nonumber\\
 -\mu n +\sum_{i,\alpha > \beta} U
n_{i\alpha} n_{i\beta},\nonumber\\
H_e=-\sum_{i,j,\alpha} t_{i,j}^{\alpha \alpha} c_{i\alpha}^+
c_{j\alpha} -\sum_{i,j,\alpha> \beta} t_{i,j}^{\alpha \beta}
c_{i\alpha}^+ c_{j\beta}- \nonumber\\
\mu n +\sum_{i,\alpha > \beta} U n_{i\alpha} n_{i\beta},
\end{eqnarray}
where $a_{i\alpha(\beta)}$ is the annihilation operator of the
$t_{2g,\alpha(\beta)}-$ the orbital electrons $(\alpha(\beta)=xy,
yz, zx)$, $c_{i\alpha(\beta)}$ is the annihilation operator of the
$e_{g,\alpha(\beta)}-$ the orbital electrons
$(\alpha(\beta)=x^2-y^2, 3z^2-r^2)$ , $\mu-$ the chemical
potential, $n-$ the holes concentration,
$t_{i,j}^{\alpha \alpha} $ is the hopping integral.   \\
Let us write the system of three equations for the Green's
functions describing the hole motion in $t_{2g}$  band. When using
random phase approximation the equations set for the Green's
functions $<<a_{{\bf r},\alpha}\mid a^+_{{\bf r},\alpha}>>$ and
$<<a_{{\bf r},\alpha}\mid a^+_{{\bf r},\beta}>>$  are being
closed. These equations have the following form:

\begin{eqnarray}\label{2}
(\omega-E^\alpha_{\bf k}) G^{\alpha\alpha}_{\bf k}+
\varepsilon_{\bf k} G^{\beta \alpha}_{\bf k}+\varepsilon_{\bf k} G^{\beta^{'}\alpha}_{\bf k}=1 \nonumber\\
\varepsilon_{\bf k}G^{\alpha\alpha}_{\bf k}+ (\omega-E^\beta_{\bf
k})
G^{\beta\alpha}_{\bf k}+\varepsilon_{\bf k}G^{\beta^{'}\alpha}_{\bf k}=0 \nonumber\\
\varepsilon_{\bf k}G^{\alpha\alpha}_{\bf k}+ \varepsilon_{\bf
k}G^{\beta \alpha}_{\bf
k}+(\omega-E^{\beta^{'}}_{\bf k})G^{\beta^{'}\alpha}_{\bf k}=0 \nonumber\\
G^{\alpha\alpha}_{\bf k}=<<a_{{\bf k},\alpha}\mid a^+_{{\bf
k},\alpha}>>; \nonumber\\
G^{\beta (\beta^{'})\alpha}_{\bf
k}=<<a_{{\bf k},\beta (\beta^{'})}\mid
a^+_{{\bf k},\alpha}>>\nonumber\\
E^\alpha_{\bf k}=\varepsilon_{\bf k}^\alpha-\mu+U(n_\beta+n_{\beta^{'}}) ,\nonumber\\
E^{\beta (\beta^{'})}_{\bf k}=\varepsilon_{\bf k}^{\beta (\beta^{'})}-\mu+U(n_\alpha+n_{\beta^{'}(\beta)} ),\nonumber\\
 \varepsilon_{\bf k}^{\alpha (\beta)}=-4t_{xy} \cos{\frac{k_x}{2}}\cos{\frac{k_y}{2}}
-4t_{xz} \cos{\frac{k_x}{2}}\cos{\frac{k_z}{2}}-\nonumber\\
4t_{zy} \cos{\frac{k_z}{2}}\cos{\frac{k_y}{2}} -2t_x
\cos{k_x}-2t_y \cos{k_y}-2t_z \cos{k_z},
\nonumber\\
\varepsilon_{\bf k}=-2t ( \cos{k_x}+ \cos{k_y}+\cos{k_z})\nonumber\\
n=n_1+n_2+n_3  \;
\end{eqnarray}
The values $\varepsilon_{\bf k}^{\alpha }$ have a different set of
hopping integrals parameters $\alpha=xy, t_{xy} >> t_{yz}, t_{xz},
\alpha=yz, t_{yz} >> t_{xz}, t_{xy}, \alpha=xz, t_{xz} >> t_{yz},
t_{xy} $.
 The equations set for the Green's functions
$<<c_{{\bf r},\alpha}\mid c^+_{{\bf r},\alpha}>>$ and $<<c_{{\bf
r},\alpha}\mid c^+_{{\bf r},\beta}>>$ describing the carriers
dynamic in the $e_g$ sub-bands  has the following form:

\begin{eqnarray}\label{3}
(\omega-A^\alpha_{\bf k}) G^{\alpha\alpha}_{\bf k}+
\varepsilon_{\bf k} G^{\beta \alpha}_{\bf k}=1 \nonumber\\
\varepsilon_{\bf k}G^{\alpha\alpha}_{\bf k}+ (\omega-A^\beta_{\bf
k})
G^{\beta\alpha}_{\bf k}=0 \nonumber\\
G^{\alpha\alpha}_{\bf k}=<<c_{{\bf k},\alpha}\mid c^+_{{\bf
k},\alpha}>>; G^{\beta \alpha}_{\bf k}=<<c_{{\bf k},\beta }\mid
c^+_{{\bf k},\alpha}>>\nonumber\\
A^{\alpha(\beta)}_{\bf k}=\varepsilon_{\bf k}^{\alpha(\beta)}-\mu+U n_{\beta(\alpha)} ,\nonumber\\
\varepsilon_{\bf k}^{\alpha (\beta)}=-2t_x^{\alpha}
\cos{k_x}-2t_y^{\alpha}
\cos{k_y}-2t_z^{\beta} \cos{k_z}\nonumber\\
\varepsilon_{\bf k}=-2t ( \cos{k_x}+ \cos{k_y}+\cos{k_z}). \nonumber\\
\end{eqnarray}
The type of $\varepsilon_{\bf k}^{\alpha (\beta)}, \alpha=x^2-y^2,
\beta=3z^2-r^2 $ reflects the symmetry of the $sp$ wave functions
hybridization of sulphur ion with the $d_{x^2-y^2}, d_{3z^2-r^2}$
wave functions .

 The solution  of Eq.(2) is reduced  to the
following cubic equation for determining of the excitation
spectrum:
\begin{eqnarray}\label{4}
\omega^3-A\omega^2+B\omega+A\varepsilon_{\bf
k}^2+2\varepsilon_{\bf k}^3-E^{xy}_{\bf k}E^{xz}_{\bf
k}E^{yz}_{\bf k}=0
\nonumber\\
A=E^{xy}_{\bf k}+E^{xz}_{\bf k}+E^{yz}_{\bf k}; \nonumber\\
B=E^{xy}_{\bf k}E^{xz}_{\bf k}+E^{xy}_{\bf k}E^{yz}_{\bf
k}+E^{xz}_{\bf k}E^{yz}_{\bf k}-3\varepsilon_{\bf k}^2.
\end{eqnarray}
The excitation spectrum in $e_g$ sub-bands is found from the
Eq.(3):
\begin{eqnarray}\label{5}
\omega_{1,2}(k)=\frac{1}{2}(A^{\alpha}_{\bf k}+A^{\beta}_{\bf k}
\pm \sqrt{(A^{\alpha}_{\bf k}-A^{\beta}_{\bf k})^2+4
\varepsilon_{\bf k}^2})
\end{eqnarray}
 The chemical potential is calculated from the self-consistent
equation for the hole concentration $n$
\begin{equation}
n=\frac{1}{N} \sum_{{\bf k},\alpha} \int d\omega  f(\omega)
\frac{1}{\pi} {\rm Im} G^{\alpha\alpha}({\bf k}\omega),
\end{equation}

where $f(\omega)=({\rm exp} (\omega/T)+1)^{-1}$. The summation
over the momentum in Brillouin zone is made using $8 \cdot 10^6$
points. Holes distribution function in $e_g$ and $t_{2g}$
sub-bands is calculated by this expression:

\begin{widetext}
\begin{eqnarray}\label{7}
N({\bf k})=\int d\omega  f(\omega) \frac{1}{\pi} {\rm Im}
G^{\alpha\alpha}({\bf k}\omega), \;  N_e^\alpha({\bf
k})=\frac{A^\alpha_{\bf k}(f(\omega_2)-f(\omega_1))+\omega_1
f(\omega_1)- \omega_2
f(\omega_2)}{\omega_1-\omega_2}\nonumber\\
N_t^\alpha({\bf k})=f(\omega_1)\frac{\varepsilon_{\bf
k}-(\omega_1-E^{\beta}_{\bf k})(\omega_1-E^{\beta^{'}}_{\bf
k})}{(\omega_1-\omega_2)(
\omega_1-\omega_3)}+f(\omega_2)\frac{\varepsilon_{\bf
k}-(\omega_2-E^{\beta}_{\bf k})(\omega_2-E^{\beta^{'}}_{\bf
k})}{(\omega_2-\omega_3)( \omega_2-\omega_1)}+
f(\omega_3)\frac{\varepsilon_{\bf k}-(\omega_3-E^{\beta}_{\bf
k})(\omega_3-E^{\beta^{'}}_{\bf k})}{(\omega_3-\omega_2)(
\omega_3-\omega_1)},
\end{eqnarray}
\end{widetext}
where $\omega_i$ are determined from Eq.4 and Eq.5 .  Holes
concentrations in sub-bands $(n_1, n_2)$ are found by minimization
of grand canonical potential of the holes gas :

\begin{eqnarray}
\frac{\partial U_g}{\partial n_1}+\frac{\partial U_g}{\partial n_2}=0 \nonumber\\
U_g=\frac{1}{N}\sum_{{\bf k},\alpha} [\omega ({\bf k})f(\omega
({\bf k}))- \nonumber\\
k_BT f(\omega ({\bf k}))\ln f(\omega ({\bf k}))]-\mu n.
\end{eqnarray}\\

The transport properties such as a conductivity can be obtained
from Kubo formula  in the limit of $d \to \infty $ \cite{Isyumov}
\begin{eqnarray}
\sigma(\omega)=\sigma_0 \sum_{\alpha} \int d\omega^{'}
I_{\alpha}(\omega^{'},\omega^{'}+\omega)
\frac{f(\omega^{'})-f(\omega^{'}+\omega)}{\omega},\nonumber\\
I_{\alpha}(\omega_1,\omega_2)=\frac{1}{\pi^2}\sum_{{\bf k}} {\rm
Im} G_{\alpha}({\bf k},\omega_1){\rm Im} G_{\alpha}({\bf
k},\omega_2).\;
\end{eqnarray}
where $\sigma_0$ is a constant defining the conductivity
dimension. In order to estimate $U$ Coulomb potential, we used the
relation between intra- and interband Coulomb parameter $U^{'}/U
\sim 0.6$ determined on basis of the first principle calculation
method for the perovskite  compounds \cite{Solovyev}. To achieve
the best agreement between the theoretical results and the
experimental data, such as the temperature dependence of
resistivity , the activation energy , the current derivative of
voltage and the optical absorption spectrum, we obtained the
values for the following parameters
$(t_x,t_y,t_z,t_{xy},t_{xz},t_{yz})$ used in Eqn.2
$(0.4,0.4,0.04,1,0.1,0.1)t_0, (0.4,0.04,0.4,0.1,1,0.1)t_0$ and
$(0.04,0.4.0.4,0.1,0.1,1)t_0, t_0=0.067 eV $ for $d_{xy}, d_{yz},
d_{zx}-$ orbital, respectively and $(t_x,t_y,t_z)$ used in Eqn.3
$(0.1,0.1,1)t_0, (1, 1, 0.1) t_0 , t_0=0.11 eV $ for $(3z^2-r^2),
(x^2-y^2) $ orbital, respectively. These bands are occupied by
holes with the average filling number $n_t=0.015, n_e=0.04$ .The
Coulomb integral between different electron orbitals  is $U=2 eV$.
The hybridization of subbands may result from the electron-phonon
interaction or interaction of different orbitals via anion with
effective hopping parameter $t=0.05 t_0$. \\

{\bf 4. Discussion }\\

Calculated optical conductivity $\sigma(\omega)$ for $e_g$ band
 reveals the optical quasigap at $\omega=0.28 eV$.
The temperature dependence of the normalized conductivity
$\sigma(\omega \to 0)$ calculated at $\omega \simeq 10^{-4} eV $
is presented in Fig.7. The conductivity realized by the carriers
in $e_g$ band decreases with the temperature increase, similar to
electrons gas in metals. The small $\sigma(T)$ value  results from
Coulomb gap $\Delta_c$ at the Fermi level .  The
$\Delta_c=E_F-\mu$ dependence  on temperature fits well the linear
$(E_F-\mu) \simeq 11 T $. The typical temperature dependence of
resistivity for semiconductors is $\rho \sim \rho_0
exp(\Delta_c/T)$ and $\rho \sim \rho_0 exp(11)$ for the $ \alpha-
MnS $ at $ T < T_N $. The resistivity is independent of
temperature, which is in good agreement with the experiment.  \\
The conductivity of $t_{2g}$ band attributes to the thermally
activated carriers. Chemical potential lies in the range of
energies between two peaks forming the effective quasigap in the
excitation spectrum, as shown in Fig.8b. As one can see from
Fig.7, the holes in $e_g$ and in $t_{2g}$ bands make the main
contribution to the conductivity   at $T < 200 K $ and at $T> 200
K $, respectively.

Calculated resistivity  $\rho=1/\sigma, \sigma=\sigma_t+\sigma_g$
is presented in Fig.9. Good agreement with the experimental data
obtained for the $\alpha-MnS$ single crystal is observed. In Fig.9
$\rho$ is normalized to the resistivity $\rho (T_1)$, where
$T_1=166 K$ corresponds to rhombohedral lattice distortion
temperature. The estimated and measured values of the activation
energy   are equal to $E_a \simeq 0.2 eV $.  Sharp resistivity
decrease at $T \sim 300 K$ arises from partial lifting of the
degeneracy of $xy, xz, zy $ subbands . Minimization of the grand
canonical potential of the holes gas $U $ with respect to the
average filling number gives the values of $n_{\alpha} \simeq n,
n_{\beta}=n_{\beta^{'}} \to 0 $ at $T< 250 K$ and
uniform filling subbands at $T > 475 K $. \\
The distribution of filling numbers $N({\bf k})$ has several
maxima   $N_{max}$ at different wave vectors ${\bf k_{m,i}}$ with
the same excitations energy. The temperature dependences of the
maximal values of $N_{max}$, ${\bf k_m}$ and the difference of the
wave vectors $\Delta {\bf k_m}={\bf k_{m1}}-{\bf k_{m2}}$
corresponding to two $N_{max}$ are shown in Fig.10. There are two
transitions associated with the change of the wave vector ${\bf
k_{m1}}$ at $T=250 K$ and  at $T=475 K$. The change of  ${\bf
k_{m2}}$ with respect to ${\bf k_{m1}}$ is observed in $(xy)$
plane at $T<300 K$ . At $T
> 475 K$ the distribution function $N({\bf k})$ has only one
maximum. Variance $D({\bf q})$ of the distribution function
$N({\bf  k})$ is calculated as $D({\bf q})=\sum_{\bf k}N({\bf k})
N({\bf k+q})-<N>^2$ . The maximal value of $D({\bf q_m})$ together
with the wave vector $q_m$ are shown in Fig.11 . The variance
minimum is reached at $T=475 K$ . The essential change of $q_m$ is
observed in temperature range $400 K < T < 470 K $. The relaxation
time $\tau$ of the current carriers  is dependent on the hole wave
number that should lead to the conductivity anisotropy. The wave
vector of the variance should be associated with the wave charge
ordering. Thus, the charge susceptibility of the holes
$\chi_c=dn/d\mu$ in $t_{2g}$ band reveals two maxima at the
temperatures $T=250 K$ and $T=475 K$ (see Fig.12). The inverse
value $1/\chi_c$ in $e_g$ band reaches minimum at $T \simeq 160 K
$ near the temperature of the transition $(T_1=166 K)$ attributed
to the rhombohedral deformation of the lattice.\\
 The resistivity value with uniform distribution of the average filling number in
 subbands $d_{xy}, d_{xz}, d_{yz}$ is more than the
resistivity of system with one partially filled sub-band at $T<
500 K$ (Fig.9). This is in qualitative agreement with the
experimental data obtained at heating of $\alpha-MnS$ single
crystal as shown in Fig.1. The temperature hysteresis $\rho(T)$
observed for $\alpha-MnS$ (Fig.1) should be caused by the
conservation of charge ordering corresponding to rhombohedral
deformation of lattice at the temperatures above $T_1=166 K$ . We
suppose that the single crystal decomposes into degenerate domains
with $n_\alpha \simeq n, n_\beta \to 0$  . During cooling from $T
\sim 500 K$ in magnetic or electric field the single crystal is
poling. To pass from one state to another
 it is necessary to overcome a potential barrier . If an external
voltage $V$ is applied  to the single crystal, carriers tunnelling
can be observed at the chemical potential surface . The density of
states (DOS) of the single holes excitations in the vicinity of
the chemical potential $(\mu-E) \sim 0.1 eV$ can be determined by
voltage differentiation of current $dI/dV$

\begin{eqnarray}
dI/dV \propto \int_{-\infty}^{\infty}d\omega g(\omega)
\frac{\partial}{\partial (eV)} f(\omega-eV) \propto g(eV).
\end{eqnarray}
Fig.2 presents the experimental data $dI/dV$, which well agree
with the estimated $g(\omega)$ dependence. The small value of the
negative differential resistivity within $\sim 5\%$ error for $V=2
V, 30
V$ reproduces a fine structure of the DOS near the chemical potential.\\
The DOS of $t_{2g}$ band has two maxima  that allow to understand
the origin of the  two maxima of the optical absorption spectra in
$\alpha - MnS $  at $\omega \simeq 2.4 eV$ and $\omega \simeq 2.9
eV$ \cite{Romanova}. They can be ascribed to the single electron
transition from the sulphur ion  to the manganese ion. The line
shape of the optical absorbtion is presented in Fig.8c and
qualitatively agrees  with the DOS of $t_{2g}$ band. The lower
band edge shows the red shift $\Delta W^{theory} \sim 0.01 eV ,
\Delta W^{exper} \sim 0.03
eV $ at temperature increase from $168 K$ to $300 K$ .  \\
Spin-orbital interaction of the holes located in the two sub-bands
$t_{2g}$ with $z-$ components of the orbital spin $L^z=\pm 1$
causes a spin splitting of the band and leads to the the
spontaneous magnetization

\begin{equation}
m=\int d\omega g(\omega) f(\omega+\frac{\lambda}{2} \sigma)-\int
d\omega g(\omega) f(\omega-\frac{\lambda}{2} \sigma),
\end{equation}
where $\sigma $ is the parameter of the  orbital ordering, which
disappears at $T \simeq 475 K$.For simplicity the temperature
dependence of $\sigma(T)$ is calculated in terms of Brillioun
function with the orbital spin $L=1$. The measured and calculated
values of the spontaneous magnetization are shown in Fig.3. The
minimum in the inverse value of the charge susceptibility at $T
\sim 250 K$ correlates with the small deviation $m^{ex}(T)$ from
estimated temperature dependence $m^{th}(T)$ . Parameter of the
spin-orbital interaction $\lambda \simeq 6 cm^{-1} $ is determined
from the magnetization calculation according to Eq.(11). The
estimated values $m(\lambda)$ fit well by the linear dependence
$m=0.061 \lambda/t_0; t_0=0.067 eV $. \\
The charge ordering induces the local deformation of the lattice
and leads to the lowering of the crystal symmetry that  can be
observed  from the electron paramagnetic resonance data. The axial
and rhombic terms of the single ion anisotropy  $D$ and $E$ are
determined from the temperature dependence of $H_{res}$ using the
general formula for the resonance shift due to crystal-field . The
expression for the effective $g-$ value for $H_{ext}$ applied
along one of the crystallographic axes was obtained in \cite{ESR}:
\begin{equation}
\frac{g^{eff}_{a,c}(T)}{g_{a,c}} \propto
1+\frac{D}{T-T_{CW}}[(3\zeta-1) \pm 3 (1+\zeta) sin(2\gamma)],
\end{equation}
where $T_{CW}$ is  the Curie-Weiss $(CW)$ temperature ;
$\zeta=E/D$ and $\gamma$ is the rotation angle of the $MnS_6$
octahedra . At $T> 200 K$  the CW law of the magnetic
susceptibility is satisfied and the data are  described by this
approach (solid lines in Fig.5), where $T_{CW}$ was kept fixed at
$475 K$, the rotation angle $\gamma=0$, $D=0.40(6) K$, the $E/D$
is the ratio $\zeta=0.016(5)$ and the $g-$ values
$g_{[100]}=1.992(8), g_{[001]}=1.984(4)$.
 Our result agrees
with $D=0.34 K $ value determined from antiferromagnetic resonance
 \cite{resonans} where  the gap in the  magnons excitation
 spectrum at  ${\bf k}=0$  was found to be equal to $\simeq 3.28 cm^{-1}$
  . The orbital ordering  gives rise  to the nonlinear
behavior $\chi(T)$ in small magnetic  fields . The effect of
strong irreversible change of the magnetization (Fig.3) versus
temperature at heating and cooling in the small magnetic field  is
explained by conservation of the degeneration of the holes in
$t_{2g}$ subbands. At cooling in the magnetic field from
temperature $T > 250 K$ this degeneration is lifted and holes
occupy  the state with the
orbital moment directed  along the external magnetic field . \\

{\bf 5. Conclusion} \\

The total conductivity of the $MnS$ is the result of motion of
holes in $e_g$ and $t_{2g}$ bands . The holes in $e_g$ band are
responsible for the temperature independent behavior  of
conductivity at low temperatures $T< T_N$ . The sharp decrease of
the resistivity at $T>200 K$ is caused by the thermal activation
of the holes in degenerate $t_{2g}$ band. The nonlinear behavior
of $\rho(1/T)$ at $T > 350 K$ and temperature hysteresis of the
conductivity at $ T< 500 K$ arise from partial lifting
degeneration of the holes in $t_{2g}$ sub-bands observed at $250 K
< T < 475 K $. The filling of two $t_{2g}$ subbands at $T\simeq
475 K$ , one of three $t_{2g}$ sub-bands at $T \simeq 250 K $ and
one of two $e_g$ sub-bands at $T \simeq 160 K$ induces the charge
instability due to the competition between  the on-site Coulomb
interaction of the holes
in the different orbitals and small hybridization of the sub-bands. \\
The localization of the charges in certain orbitals causes the
spin polarization of the current carriers and leads to the weak
ferromagnetism. Orbital ordering  gives rise
 to the anisotropy of the g- values.  \\

We thank to prof.  S. G. Ovchinikov for useful discussion.\\
 This
work is supported by the Russian Basic Research Foundation $(RFFI
- BRFFI$ , project no. $04 - 02 - 81018 Bel 2004_a)$

\newpage

Fig.1 The temperature dependence of the resistivity measured at
two cycles of heating and cooling: the first (1) and the second
(2) cycles.

Fig.2 The voltage derivative of current $dI/dV (1)$ and density of
state near the chemical potential $g(\omega) (2)$ at $T=280 K (a);
550 K (b)$ .

Fig.3 The temperature dependence of the residual magnetization on
the one manganese ion along $[111](a)$ measured in zero field cold
(ZFC)  and at $H=200 Oe  $ (FC). Estimated (1) by Eqn.(11) and
measured (2) magnetization  along $[001]$ versus temperature (b) .

Fig.4 The magnetization vs. magnetic field at the different
temperatures .

Fig.5 The temperature dependence of the effective $g-$ value
$g_{eff}(T)(a) $ and $\Delta H $ (1),$ \Delta H_{[001]}-\Delta
H_{[100]}$ (2) (b) for $H$ parallel to the crystallographic axis .
Solid lines represent the fits using Eqn.(12) for  $g_{eff}$ .

Fig.6 Schematic drawing of the $t_{2g} $ holes moving on the sites
and density of states of single electron excitations in $t_{2g} $
(solid) and $e_g$ (dotted) lower Hubbard band (LHB) and upper
Hubbard band (UHB) with arrangement of Fermi level $(E_F)$ and
chemical potential \cite{Taperro}.

Fig.7 The  temperature dependence of the hole conductivity  in
$t_{2g} (1)$ and in $e_g (2)$ bands  normalized to the constant
$\sigma_0$ defining the dimension of conductivity  .

Fig.8 The density of states $g(\omega)$ of single particle
excitations in $e_g$ band at $T=100 K (1), 400 K(2) (a)$ and in
$t_{2g}$ band at $T=200 K (1), 700 K(2) (b)$ which energy is
normalized to hopping parameter $t$. The optical absorbtion
spectrum measured in \cite{Romanova} $(solid )$ line at $T=170 K$
and calculated density of states of holes  in $t_{2g}$ band
$(dotted)$.

Fig.9 The temperature dependence of resistivity measured $(1)$ and
calculated  for $ n_{\alpha}=n, n_{\beta}=0 (2),
n_{1}=n_{2}=n_3=n/3 (3) $ normalized to the resistivity value at
$T_1=166 K$.

Fig.10 The distribution function maximum of holes filling numbers
$N_{max}({\bf k})/N_{max}(T_1) (a)$ at wave vector $(k_x,k_y,k_z)
(b)$ in the range of temperatures $(160 - 245) K (1), (250-475) K
(2), (480-700 ) K (3) $. The difference of waves vectors $\Delta
{\bf k}={\bf k_1- k_2} (c) $ corresponding to $N_{max,1}({\bf
k_1})$ and $N_{max,2}({\bf k_2})$ satisfying to relation
$(1-N_{max,2}({\bf k_2})/ N_{max,1}({\bf k_1})) \le 0.005$  .

Fig.11 The temperature dependencies of the  variance maximal value
  $D(\bf q_m)$ normalized to the
$D(\bf q_m)$  at $T=150 K$
 $(a)$ and the wave vector $ (q_x,q_y,q_z)$
corresponding to the largest $D({\bf q_m}) (b) $ , where $T=100
K(1), 500 K(2)$ .

Fig.12 The inverse value of the charge susceptibility taken
 from minimal value for $t_{2g}(1)$ and $e_g (2)$ bands
versus temperature.

% The Appendices part is started with the command \appendix;
% appendix sections are then done as normal sections
% \appendix

% \section{}
% \label{}


\begin{thebibliography}{11}


\bibitem{canted} M.B. Salamon  and Marcelo Jaime ,
Rev. Mod. Phys.  {\bf 73} 583 (2001).

\bibitem{Manganit} J.S. Zhou  and J.B. Goodenough   ,
Phys. Rev. B  {\bf 68} 144406 (2003).

\bibitem{magnet_res}  G.A. Petrakovskii , L.I. Ryabinkina , G.M.
Abramova, A. D. Balaev, D. A. Balaev, A. F. Bovina
  ,  Pis'ma Zh. Eksp. Teor. Fiz.  {\bf 72}
99 (2000) .

\bibitem{resist} L.I. Ryabinkina , G.M. Abramova, O. B. Romanova,
N.I. Kiselev, D. A. Velikanov, A. F. Bovina, Proceeding of second
International Symposium. Big Sochi, Russia, OMA-II, 72 (2002).


\bibitem{Heikens} H.H. Heikens , C.F. Bruggen  and C. Haas ,
J. Phys. Chem. Solids  {\bf 39} 833 (1978) .

\bibitem{Taperro} R. Tappero , P. Wolfers  and A. Lichanot,
 Chem. Phys. Lett.  {\bf 335} 449 (2001).



\bibitem{Isyumov} Yu A. Izyumov  and Yu. N. Skryabin ,
UFN  {\bf 171} 121 (2001).

\bibitem{Solovyev}  I. Solovyev , N. Hamada , K. Terakura,
  Phys. Rev. B  {\bf 53} 7158 (1996) .




\bibitem{Romanova} O.B. Romanova , G.M. Abramova , L.I. Ryabinkina , V.V.
Markov,   Phys. Met. Metallog.  {\bf 93} 85 (2002) .

\bibitem{ESR} J. Deisenhofer , B.I. Kochelaev , E. Shilova ,
A.M. Balbashov ,A.   Loidl, vonNidda,H.A.Krug ,
   Phys. Rev. B {\bf 68} 214427 (2003).

\bibitem{resonans} C.H. Perry ,E. Anastassakis , J. Sokoloff,
   J. Pure Appl. Phys.  {\bf 9} 930 (1971).



\end{thebibliography}
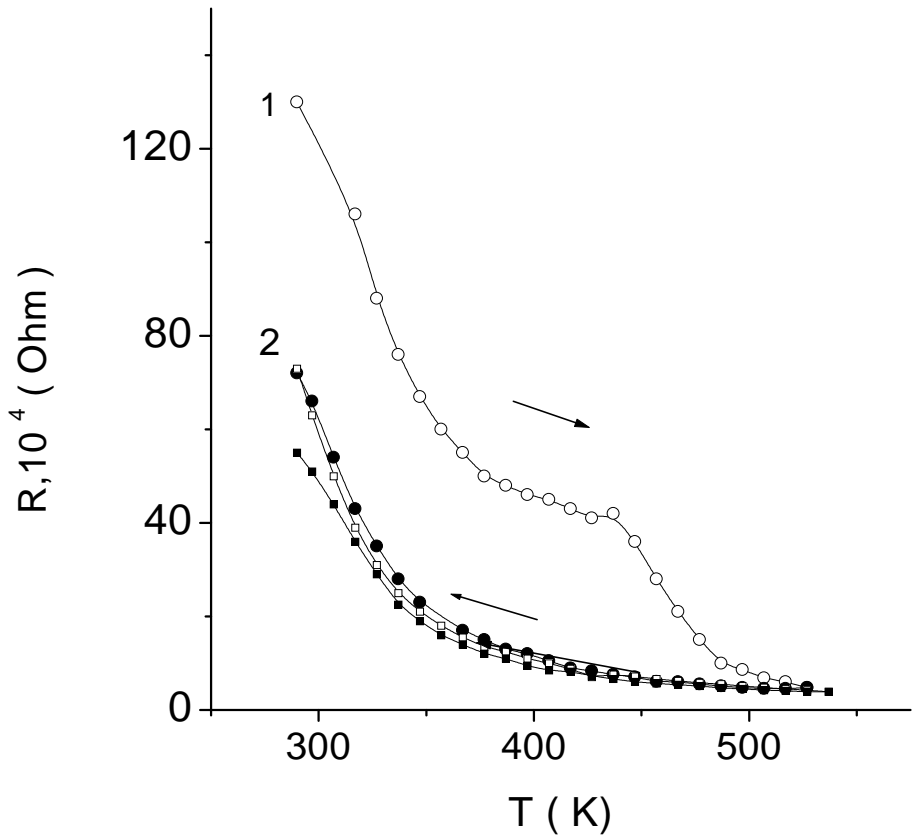
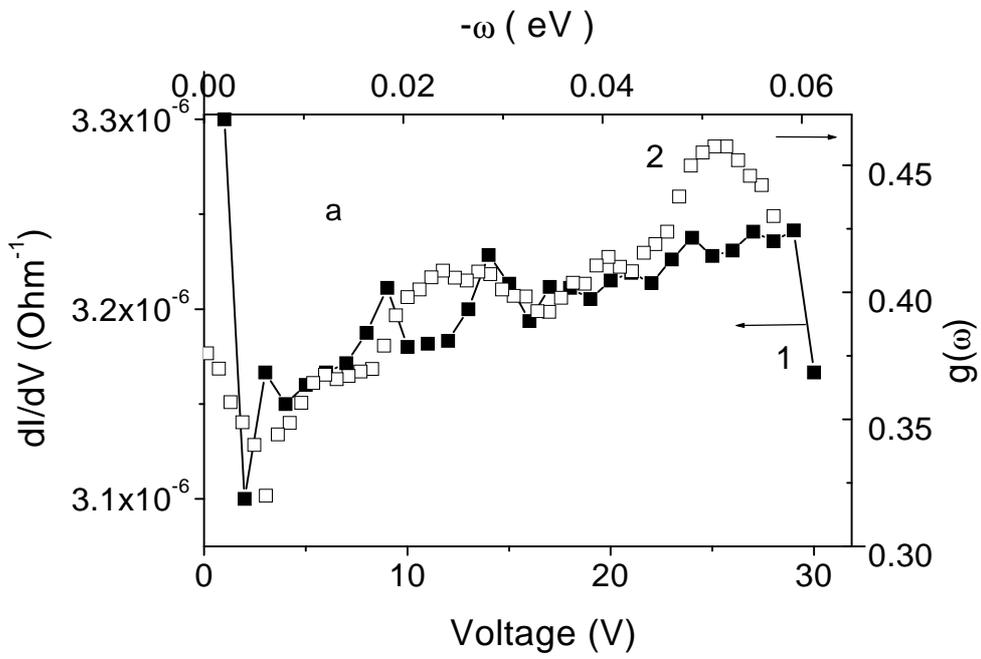
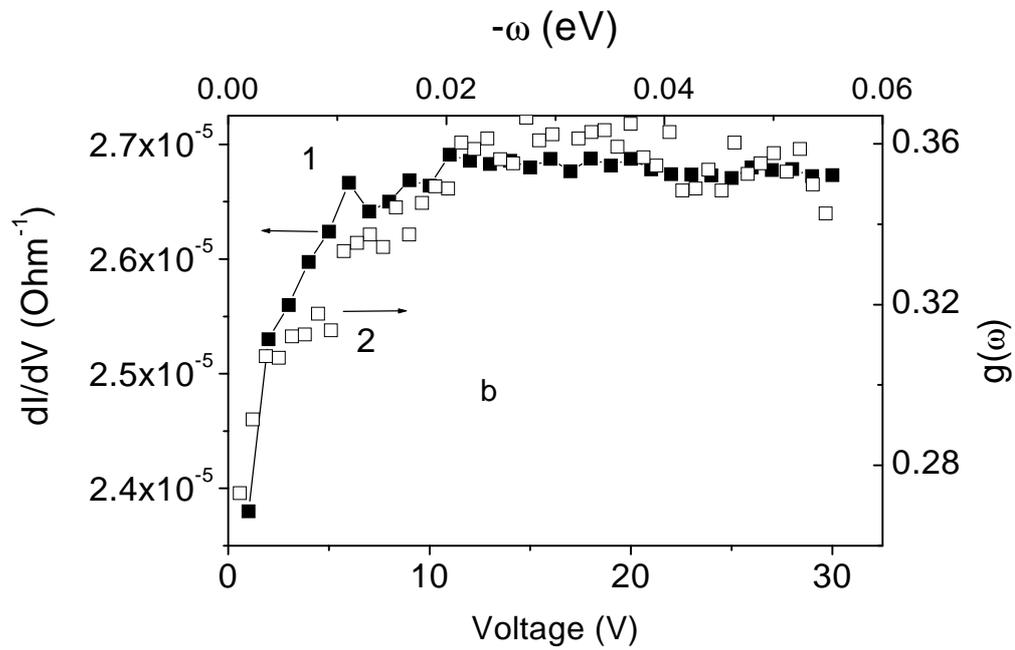
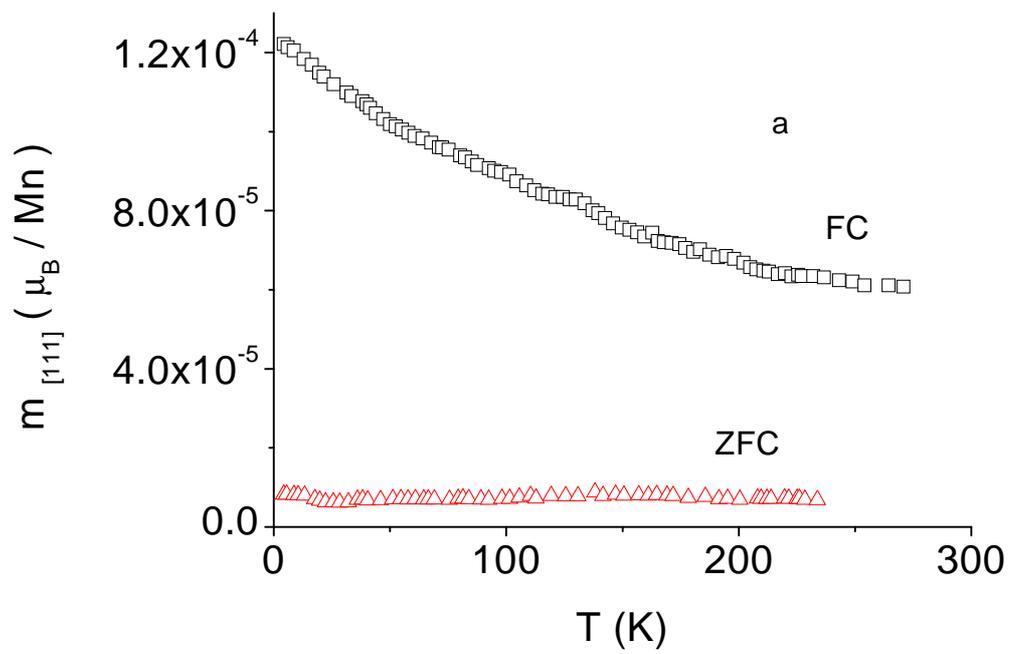
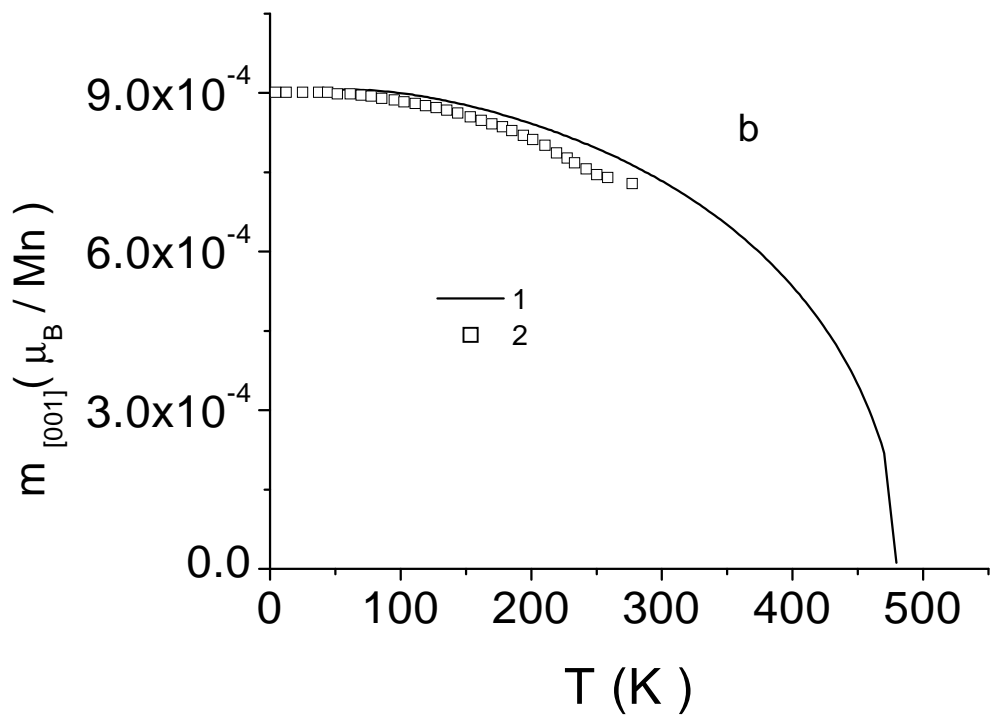
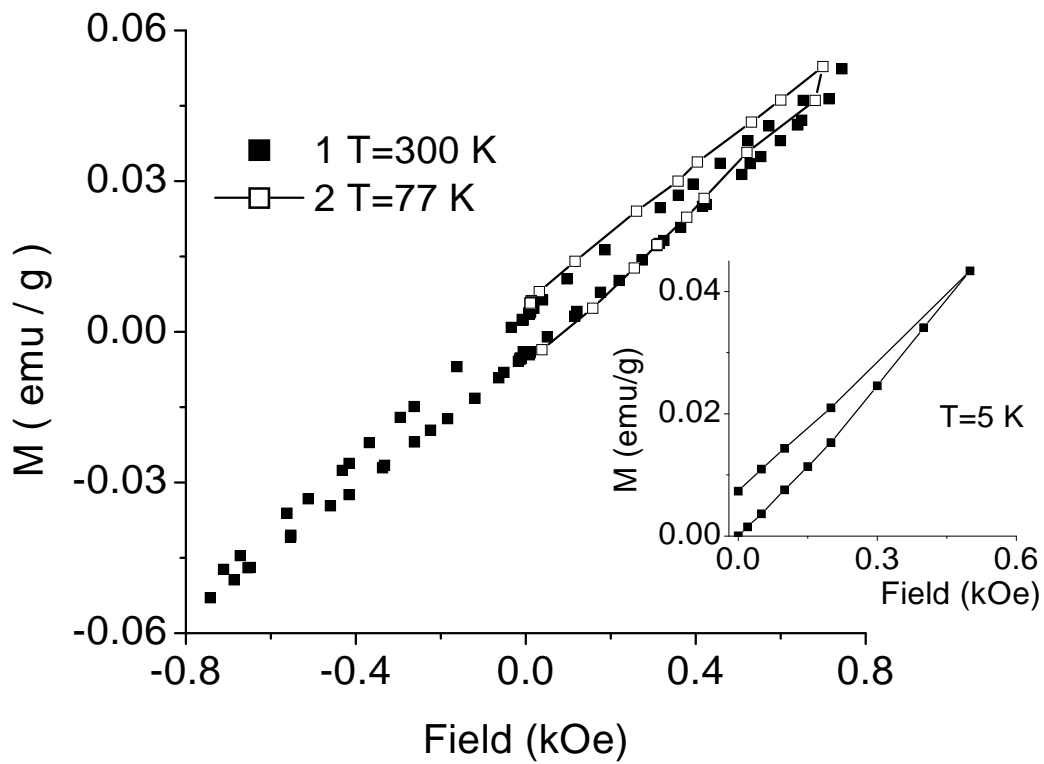
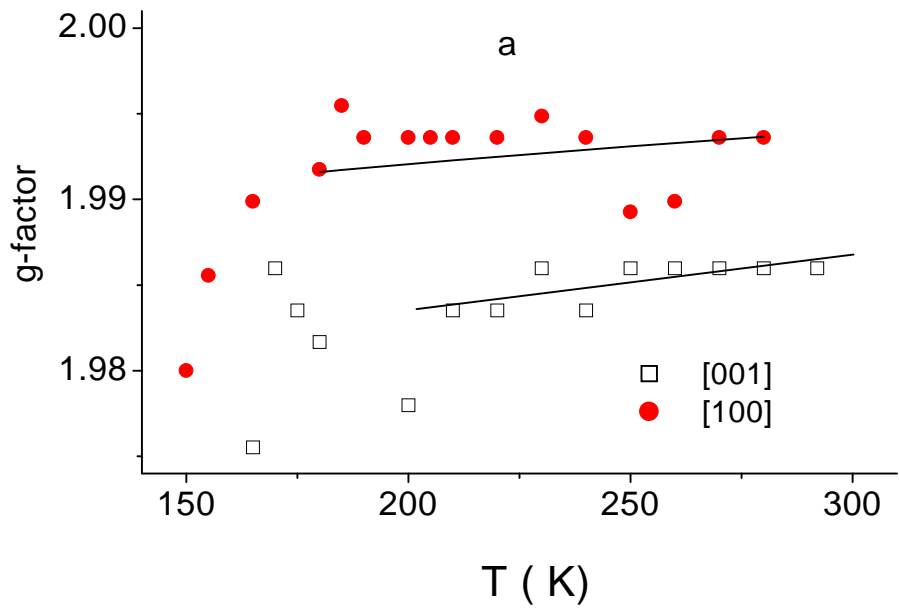
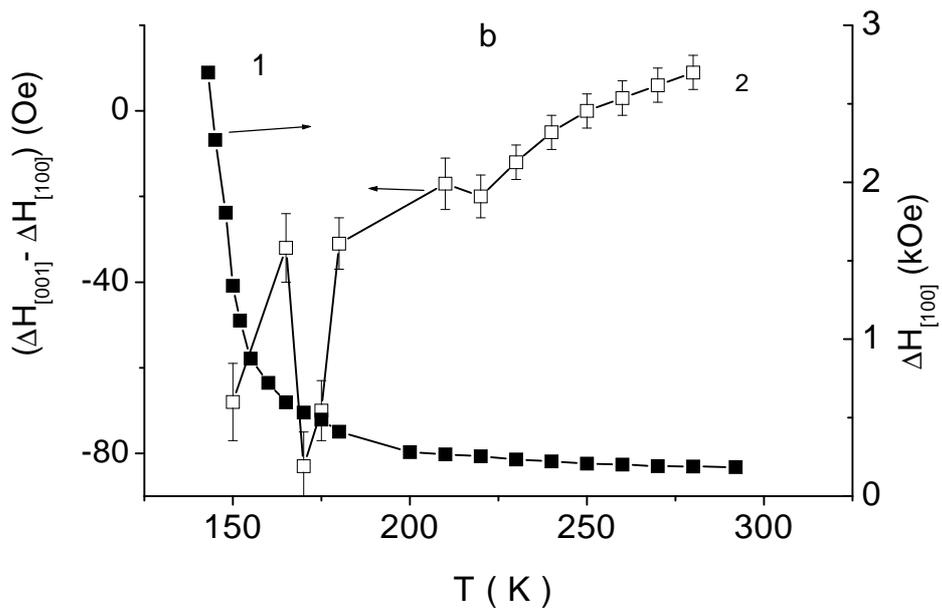
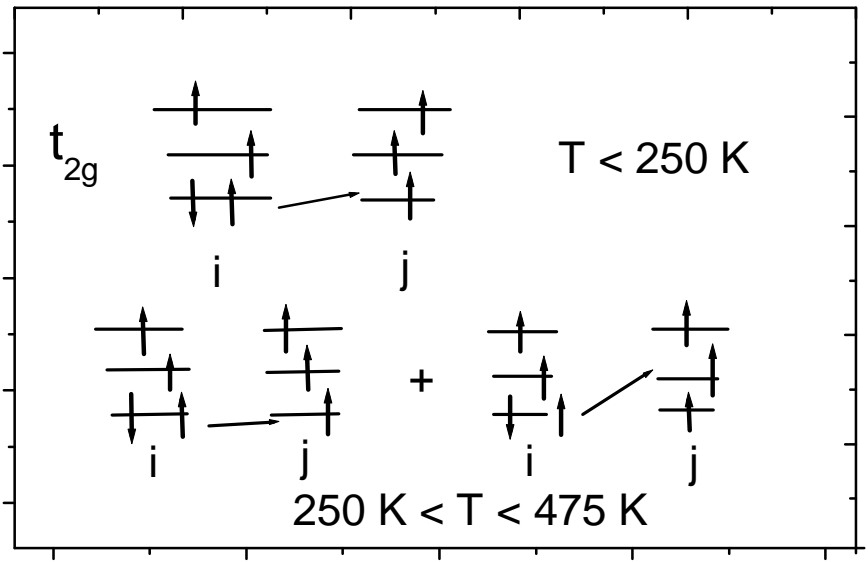
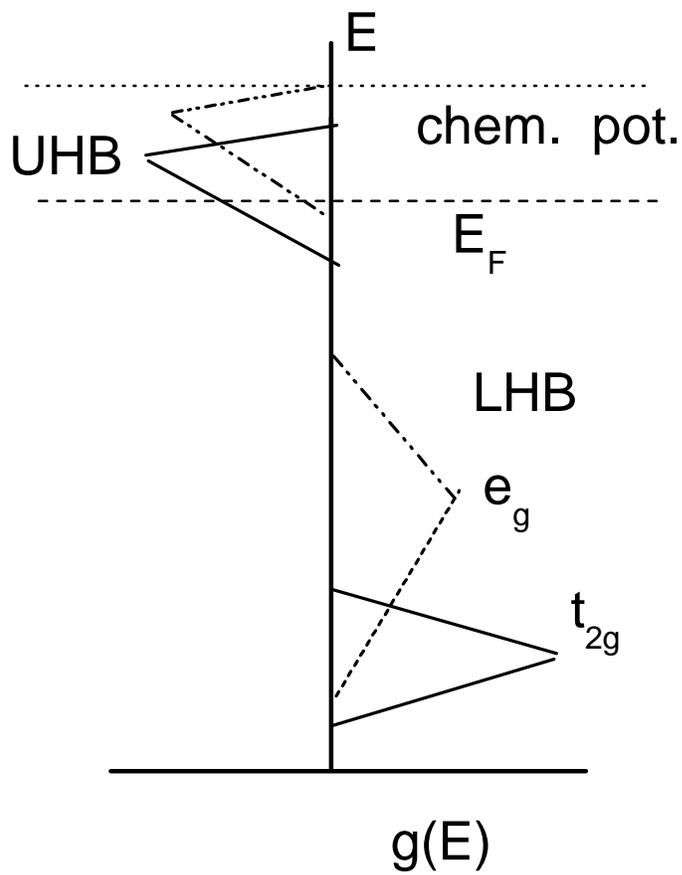
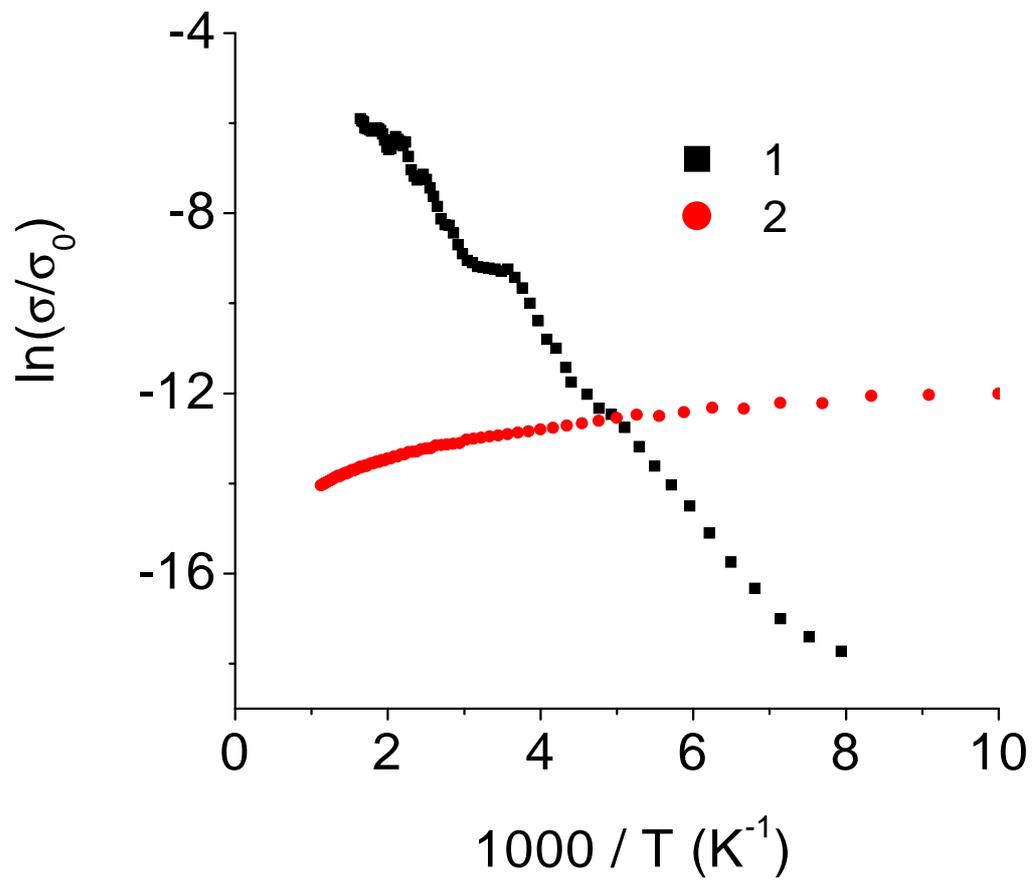
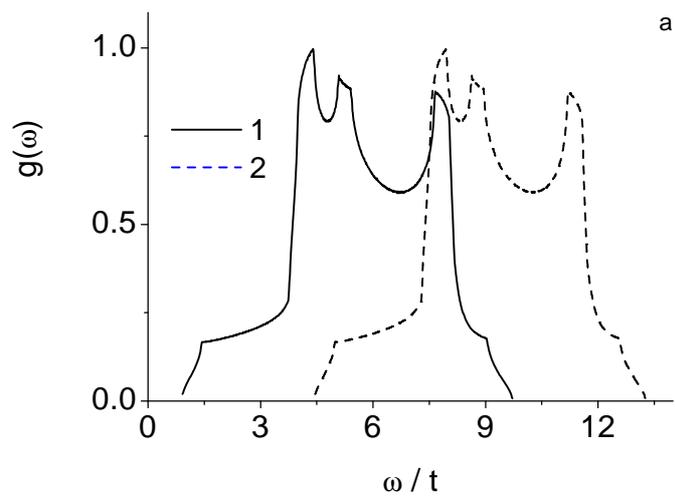
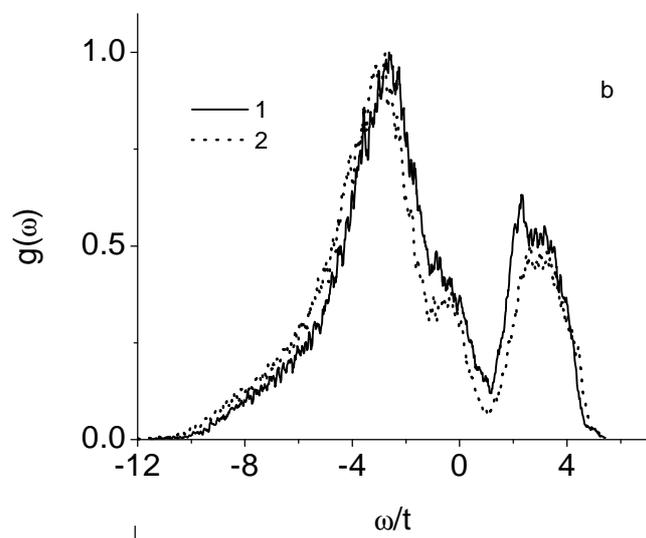
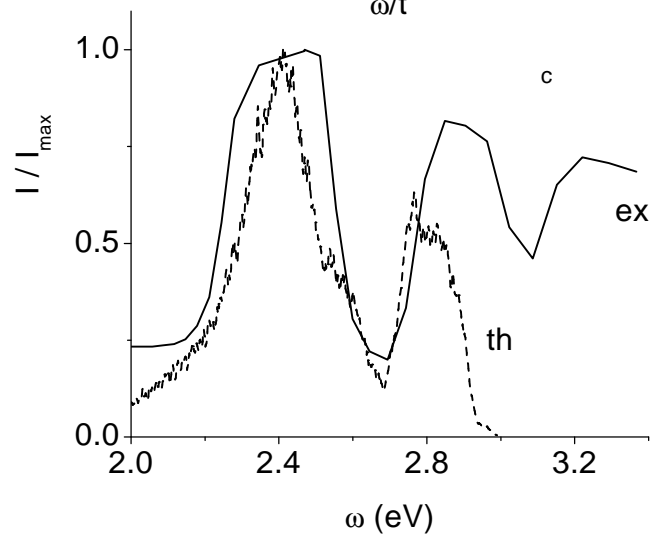
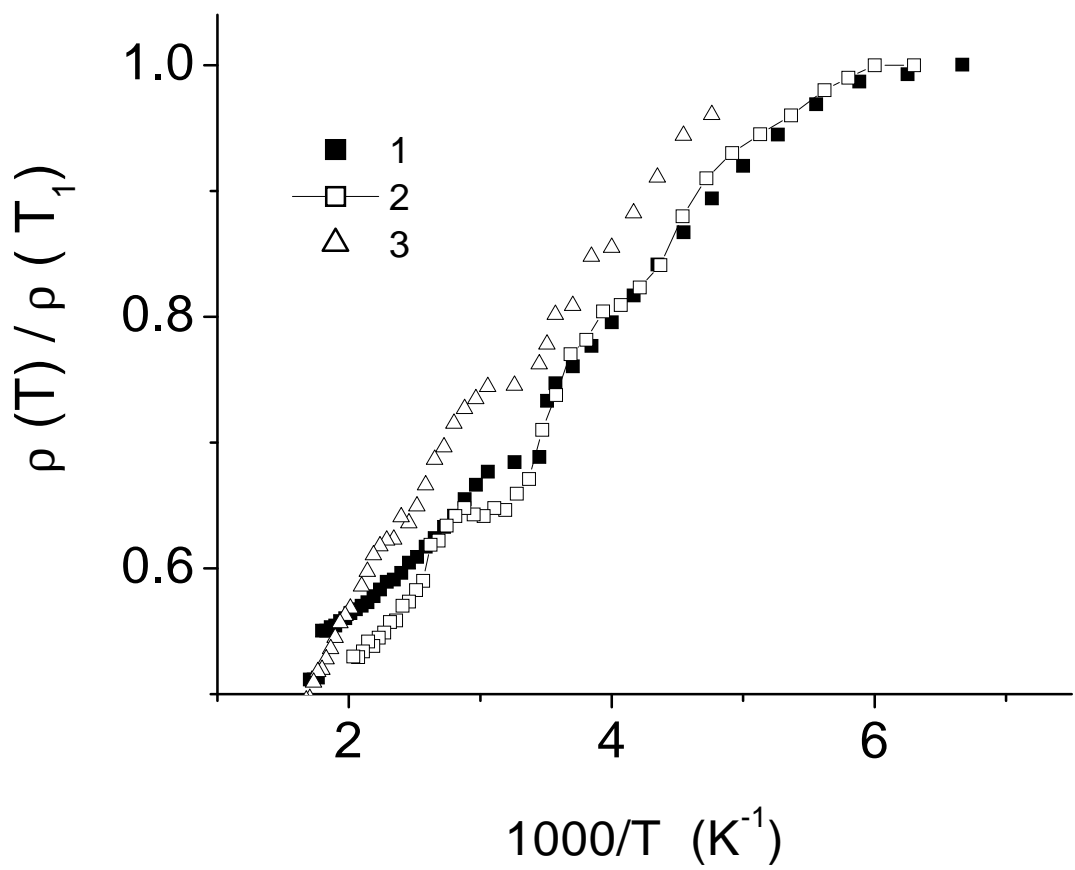
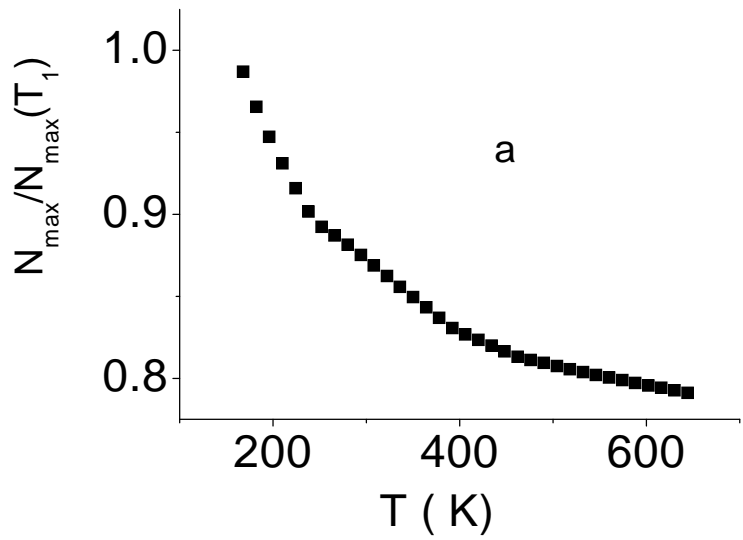
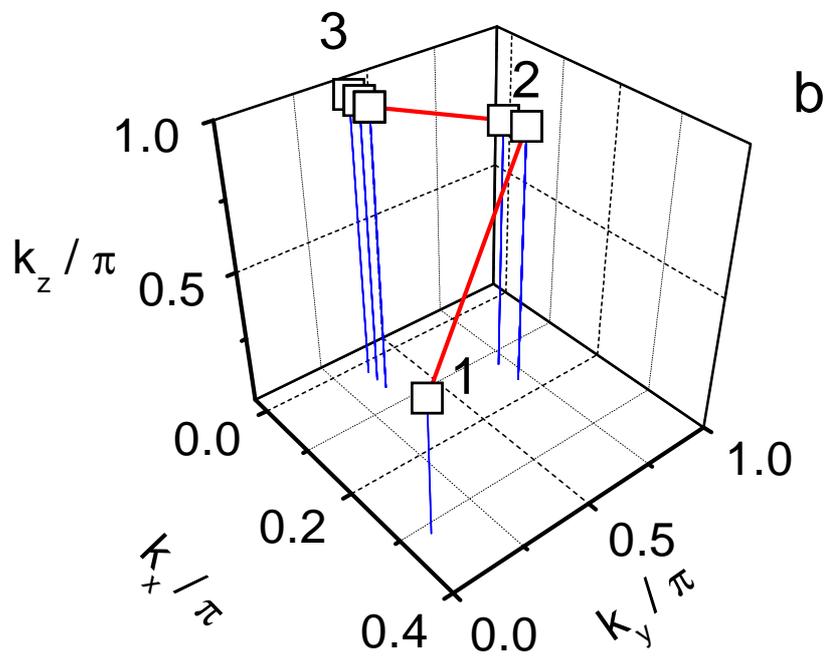
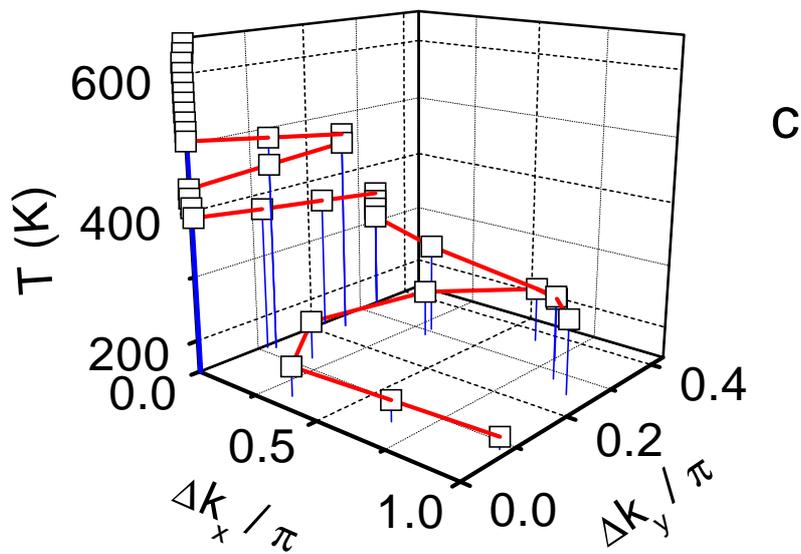
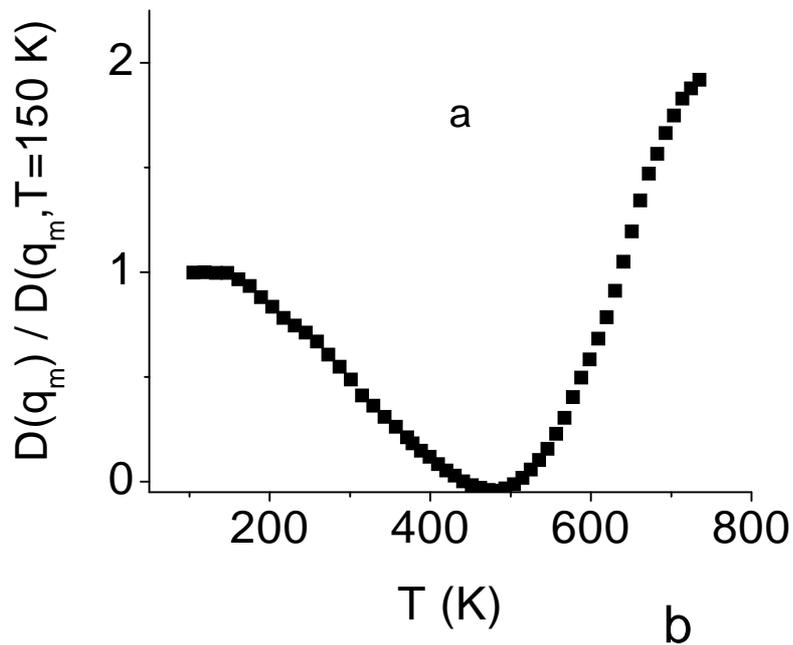
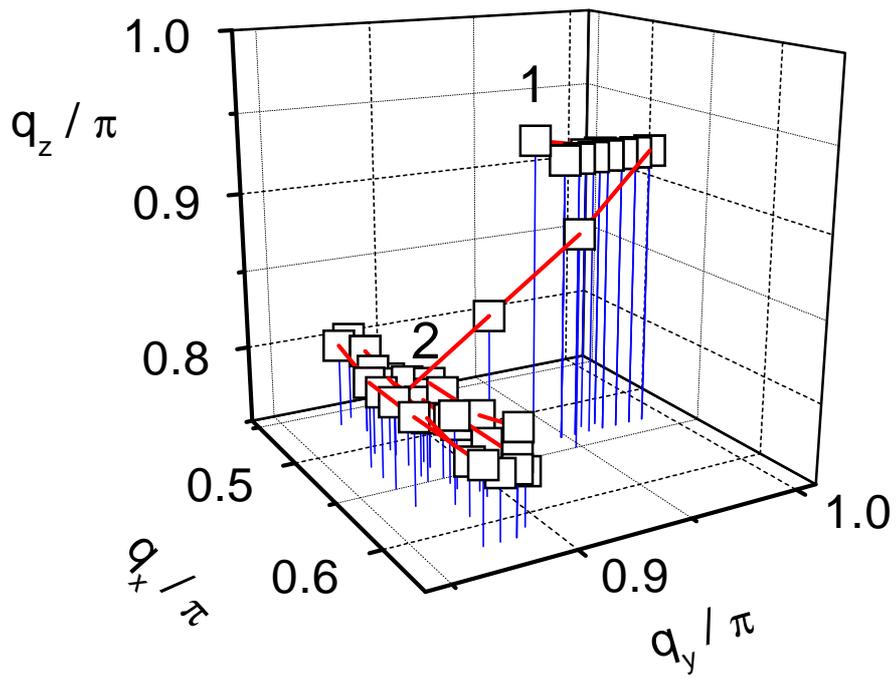
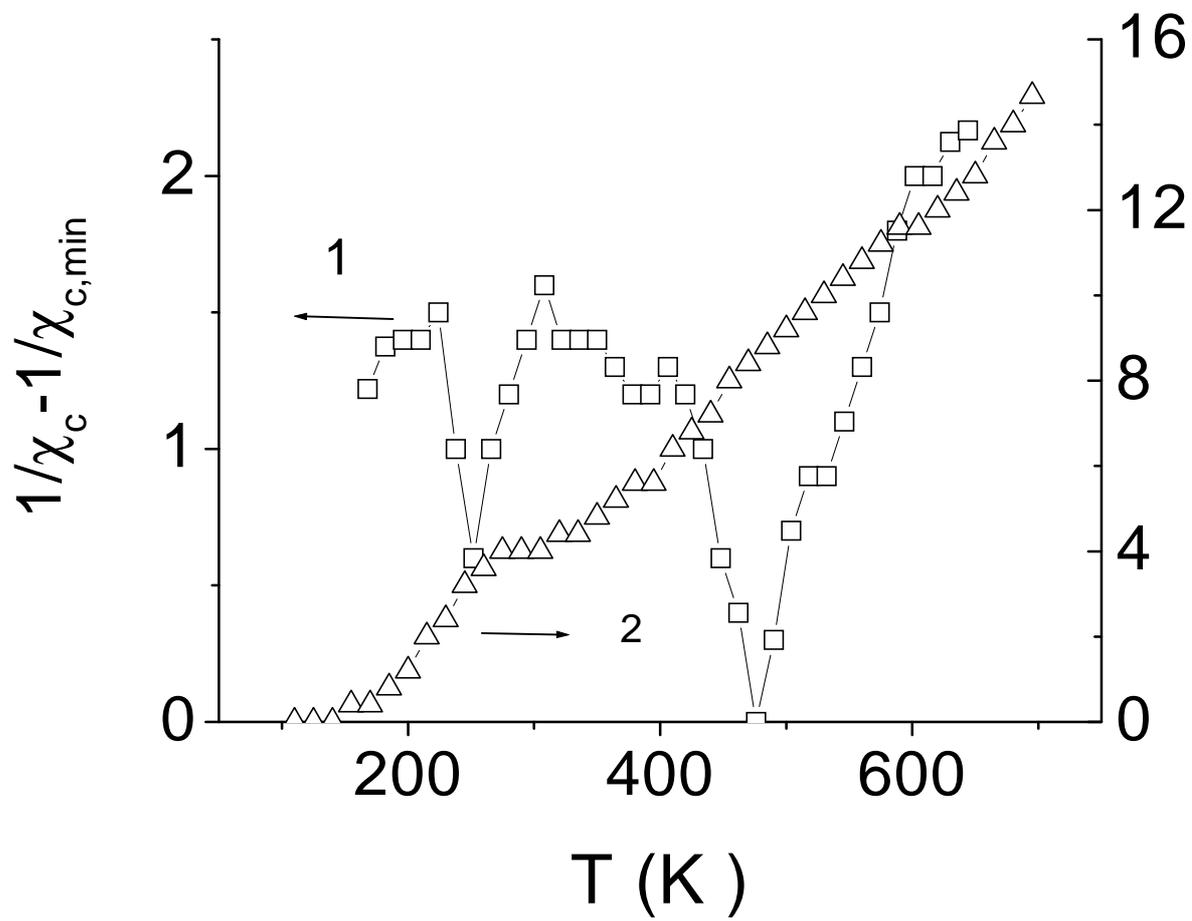
\end{document}